\documentclass[a4paper,11pt]{article}
\pdfoutput=1
\usepackage{jheppub} 

\usepackage{amsmath,amsfonts,amssymb,amsthm}
\usepackage{url}
\usepackage{physics}
\usepackage{hyperref}
\usepackage{comment}

\numberwithin{equation}{section}

\newcommand{\fft}[2]{\frac{#1}{#2}}

\newcommand{\nn}{\nonumber}

\preprint{LITP-26-17, USTC-ICTS/PCFT-26-54}

\title{\boldmath BMPV black holes in higher-derivative supergravity}

\author[a]{Yide Cai,}
\author[b]{Sabarenath Jayaprakash,}
\author[b]{James T. Liu,}
\author[c,d]{Yi Pang,}
\author[c]{Robert J. Saskowski}
\emailAdd{caiyi@umich.edu}
\emailAdd{sabare@umich.edu}
\emailAdd{jimliu@umich.edu}
\emailAdd{pangyi1@tju.edu.cn}
\emailAdd{robert\_saskowski@tju.edu.cn}

\affiliation[a]{Independent Researcher, Shanghai, China}
\affiliation[b]{Leinweber Institute for Theoretical Physics, 
University of Michigan, Ann Arbor, MI 48109, USA}
\affiliation[c]{Center for Joint Quantum Studies and Department of Physics, School of Science,\\ Tianjin University, Tianjin 300350, China}
\affiliation[d]{Peng Huanwu Center for Fundamental Theory, Hefei, Anhui 230026, China}

\abstract{There are five independent four-derivative superinvariants for the five-dimensional STU model, of which two are vector invariants that do not involve curvature tensors. We construct the corrected BMPV black hole directly in five dimensions with this set of general four-derivative couplings, a computation made possible with the help of AI. We find that the vector invariants do not correct the solution or the entropy, while the entropy associated with the heterotic corrections agrees with recent results.}

\date{\today}

\begin{document}
\maketitle
\flushbottom

\section{Introduction}
\label{sec:intro}
One of the fundamental problems in theoretical physics is the construction of a UV-complete theory of quantum gravity, which should, among other goals, provide a precise accounting of black hole entropy in terms of fundamental microstates. This has proven to be a challenging task, and we have very few examples of working quantum gravity theories. Nevertheless, string/M-theory is a shining example of a consistent theory of quantum gravity, and the microscopic entropy accounting has been successfully performed for the Strominger-Vafa black hole~\cite{Strominger:1996sh} and generalizations thereof~\cite{Breckenridge:1996is,Maldacena:1997de}. 
        
At low energies, any quantum (gravity) theory may be expanded as an effective field theory, consisting of an infinite tower of higher-derivative corrections. Such EFT expansions are best understood in the string-theoretic case, and for the heterotic string in particular, for which the expansion is known precisely to order $\alpha'^3$~\cite{Bergshoeff:1988nn,Bergshoeff:1989de}. Although, in principle, the higher-derivative action receives corrections from both UV and quantum effects, the heterotic action only receives string loop corrections starting at order $\alpha'^3$. Moreover, at order $\alpha'$, the corrected action alone is sufficient to extract the four-derivative thermodynamics via the Reall-Santos procedure~\cite{Reall:2019sah}, but obtaining the corrected solution itself is a much more difficult endeavor.

Here, we focus on the BMPV black hole~\cite{Breckenridge:1996is}, which is the rotating generalization of the Strominger-Vafa solution~\cite{Strominger:1996sh}. The four-derivative BMPV solution was recently obtained by Ruip\'erez in Ref.~\cite{Ruiperez:2026rni}, which obtained the higher-derivative corrected solution by uplifting on a torus to ten dimensions and then solving the Bianchi identity for the three-form flux and the equations of motion. Ref.~\cite{Ruiperez:2026rni} further computed the Wald entropy of the solution, resolving a puzzle of conflicting results for the entropy. Nevertheless, we are ultimately interested in the five-dimensional solution, and, as has been discussed in~\cite{Jayaprakash:2024xlr,Cai:2025yyv}, there are nontrivial field redefinitions involved in getting from the ten-dimensional solution to the five-dimensional solution when higher-derivative corrections are involved. The purpose of the present work is to bring the solution back down to five dimensions.

As was recently shown in~\cite{Cai:2026qrr}, there are five independent four-derivative superinvariants for the five-dimensional STU model. In an appropriate basis, three of these correspond to the Weyl-squared invariants of~\cite{Hanaki:2006pj}, and the other two are new $F^4$-type vector invariants that do not involve curvature tensors. One combination of these superinvariants uplifts to the Bergshoeff-de Roo action~\cite{Bergshoeff:1988nn,Bergshoeff:1989de} of heterotic supergravity in 10D. The other four superinvariants do not uplift to 10D \cite{Chang:2022urm} but can be lifted to 6D as combinations of the Riemann squared invariant \cite{Bergshoeff:1986vy} and the Gauss-Bonnet invariant \cite{Novak:2017wqc}. Our main result is the four-derivative corrected solutions corresponding to these five invariants and the entropies thereof. Just as for the static case~\cite{Cai:2026qrr}, we find that the vector invariants do not modify the BMPV solution. For the Weyl-squared invariants, rather than field redefining the ten-dimensional solution, we construct the corrected solution directly in five dimensions with the help of GPT-5.6 Sol, along with human verification of the solution.

\section{The four-derivative corrected BMPV black hole}

At leading order, the bosonic part of the two-derivative STU Lagrangian is given by
\begin{align}
    e^{-1}\mathcal L_{\partial^2}&=R-\fft12(\partial_\mu\varphi_1)^2-\fft12(\partial_\mu\varphi_2)^2-\fft1{4(X^1)^2}(F_{\mu\nu}^1)^2-\fft1{4(X^2)^2}(F_{\mu\nu}^2)^2-\fft1{4(X^3)^2}(F_{\mu\nu}^3)^2\nn\\
    &\quad+\fft14\epsilon^{\mu\nu\rho\sigma\lambda}F_{\mu\nu}^1F_{\rho\sigma}^2A_\lambda^3,
\end{align}
where the constrained scalars $X^I$ may be expressed in terms of unconstrained scalars $\varphi_i$ as
\begin{equation}
    X^1=e^{-\varphi_1/\sqrt6-\varphi_2/\sqrt2},
    \qquad X^2=e^{-\varphi_1/\sqrt6+\varphi_2/\sqrt2},
    \qquad  X^3=e^{2\varphi_1/\sqrt6}.
\end{equation}
In these conventions, the three-charge rotating BMPV black hole solution takes the form~\cite{Breckenridge:1996is}
\begin{align}
    \dd s^2&=-\fft1{\mathcal H^{2/3}}(\dd t+\omega)^2+\mathcal H^{1/3}\left(\dd r^2+\fft{r^2}4\left((\sigma_L^1)^2+(\sigma_L^2)^2+(\sigma_L^3)^2\right)\right),\nn\\
    A^I&=\fft1{H_I}(\dd t+\omega),\qquad X^I=\fft{\mathcal H^{1/3}}{H_I},
\label{eq:2dsoln}
\end{align}
where
\begin{equation}
    H_I=1+\fft{q_I}{r^2},\qquad \mathcal H=H_1H_2H_3,\qquad\omega=\fft{a}{2r^2}\sigma_L^3,
\end{equation}
and the left-invariant one-forms are given by
\begin{align}
    \sigma_L^1&=\sin\psi \,\dd\theta-\cos\psi\sin\theta \,\dd\phi,\nn\\
    \sigma_L^2&=\cos\psi \,\dd\theta+\sin\psi\sin\theta \,\dd\phi,\nn\\
    \sigma_L^3&=\dd\psi+\cos\theta \,\dd\phi.
\end{align}
This BPS black hole is parametrized by three charges, $q_1$, $q_2$ and $q_3$, and the rotation parameter $a$.

For the four-derivative corrections, we use the supersymmetrized Weyl-squared coupling constructed in the off-shell $\mathcal N=2$ supergravity approach of \cite{Hanaki:2006pj} specialized to the STU model.  These corrections are parametrized by three constants $\lambda_I$ with $I=1,2,3$, and in our field redefinition frame the full Lagrangian takes the form
\begin{equation}
    e^{-1}\mathcal L=e^{-1}\mathcal L_{\partial^2}+\lambda_IX^IR_{\mu\nu\rho\sigma}^2+D_{IJ}R^{\mu\nu\rho\sigma}F_{\mu\nu}^IF_{\rho\sigma}^J+\cdots+\fft12\lambda_I\epsilon^{\mu\nu\rho\sigma\lambda}R_{\mu\nu\alpha\beta}R_{\rho\sigma}{}^{\alpha\beta}A_\lambda^I,
\label{eq:4lagpart}
\end{equation}
where
\begin{equation}
    D_{IJ}=\fft{\lambda_I}{X^J}-\fft{\lambda_MX^M}{2X^IX^J}.
\end{equation}
The complete form of the four-derivative Lagrangian is given in \cite{Cassani:2024tvk,Cai:2026qrr}.

Note that since the STU model is symmetric under permutations of 1, 2, 3, it suffices to obtain the higher-derivative solution with only $\lambda_3$ non-vanishing.  The complete solution is then obtained at first order in $\alpha'$ by linear superposition of the permuted corrections.  In order to obtain the $ \lambda_3$-corrected solution, we make the ansatz
\begin{align}
    \dd s^2&=-e^{2f(r)}(\dd t+\omega(r)\sigma_L^3)^2+e^{2g(r)}\dd r^2+e^{2h(r)}\fft{r^2}4\left((\sigma_L^1)^2+(\sigma_L^2)^2+e^{2k(r)}(\sigma_L^3)^2\right),\nn\\
    A^I&=u_I(r)\dd t+v_I(r)\sigma_L^3,
\end{align}
and expand each function as $\Phi(r)=\Phi_0(r)+\lambda_3\delta\Phi(r)$, where the leading order quantities are obtained from (\ref{eq:2dsoln}).  This ansatz, along with the explicit form of the supersymmetrized Weyl-squared corrected STU Lagrangian \cite{Hanaki:2006pj,Cassani:2024tvk,Cai:2026qrr} was fed into GPT-5.6 Sol, which in turn used Mathematica for computations.  The leading-order BMPV solution served as the initial seed, and GPT-5.6 Sol was able to obtain the corrected solution in the gauge $\delta h=0$.

To match the static four-derivative corrected solution obtained in \cite{Cai:2026qrr}, we performed a coordinate transformation, $r\to\bar r(r)$, to transform to the gauge where $\delta\varphi_2=0$.  This choice is motivated by the $I=1,2$ interchange symmetry when $\lambda_1=\lambda_2=0$.  After permuting $\lambda_1$, $\lambda_2$, and $\lambda_3$ and superposing, the resulting four-derivative corrected solution takes the form
\begin{align}
    \dd s^2&=-e^{2f}(\dd t+\omega)^2+e^{2g}\dd r^2+e^{2h}\fft{r^2}4\left((\sigma_L^1)^2+(\sigma_L^2)^2+(\sigma_L^3)^2\right),\nn\\
    X^I &= \dfrac{\mathcal H^{1/3}}{H_I} \left(1 +8\left( \dfrac{(\partial \log\mathcal H)^2}{72}-\fft{a^2}{3r^8\mathcal H}\right)\sum_J\fft{\lambda_J}{H_J}(1-3\delta^I_J)\right), \nn \\
    A^I &= \dfrac{1}{H_I}\Biggl[\left(1 + 2\sum_J\fft{\lambda_J}{H_J}\left( \dfrac{\partial \log\mathcal H\partial\log(\mathcal H/H_J)}{3}-\fft{8a^2}{r^8\mathcal H}-\delta^I_J\left(\fft{(\partial\log\mathcal H)^2}3-\fft{10a^2}{r^8\mathcal H}\right) \right)\right) (\dd t+\omega)\nn\\
    &\kern4em+2\sum_J\fft{\lambda_J}{H_J}\left(\fft{7\partial\log\mathcal H+15\partial\log H_J}{3r}-\delta^I_J\fft{\partial\log\mathcal H}r\right)\omega\Biggr],
\label{eq: BMPV alpha'}
\end{align}
where $\partial$ refers to the derivative $\dd/\dd r$.  The metric functions are
\begin{align}
    e^{2f}&=\fft1{\mathcal H^{2/3}}\qty(1-2\sum_I\fft{\lambda_I}{H_I}\left(\fft{7(\partial \log(H_J/H_K))^2-2\partial\log(\mathcal H/H_I)\partial\log H_I+19(\partial\log H_I)^2}{9}+\fft{24a^2}{r^8\mathcal H}\right)),\nn\\
    e^{2g}&=\mathcal H^{1/3}\left(1-16\sum_I\fft{\lambda_I}{H_I}\left(\fft{\partial \log H_J\partial\log H_K+\partial\log(\mathcal H/H_I)\partial\log H_I}{9}-\fft{a^2}{r^8\mathcal H}\right)\right),\nn\\
    e^{2h}&=\mathcal H^{1/3}\left(1+4\sum_I\fft{\lambda_I}{H_I}\left(\fft{(\partial\log(H_J/H_K))^2+\partial\log(\mathcal H/H_I)\partial\log H_I+4(\partial\log H_I)^2}{9}+\fft{4a^2}{r^8\mathcal H}\right)\right),\nn\\
    \omega&=\fft{a}{2r^2}\left(1-4\sum_I\fft{\lambda_I}{H_I}\fft{\partial\log\mathcal H+3\partial\log H_I}{r}\right)\sigma_L^3,
\end{align}
where $(I,J,K)$ is an even permutation of 1, 2, 3.  This reduces to the static solution in \cite{Cai:2026qrr} in the limit $a=0$.

In the static case, it was shown that the STU black hole is unaffected by the presence of any additional vector multiplet superinvariants, and this continues to hold for the BMPV solution.  The key observation is the leading-order relation between vectors and scalars
\begin{equation}
    \fft{F^I}{X^I}-\fft{F^J}{X^J}=-\mathcal H^{-1/3}(\dd t+\omega)\wedge \dd\log(X^I/X^J),
\end{equation}
which generalizes the static case to include rotation.  This balance ensures that the vector multiplet invariants vanish in the Lagrangian and that the $\mathcal O(\lambda_I)$ sources also vanish in the corrected equations of motion.  This further supports the idea that the vector multiplet invariants do not modify the nature nor the thermodynamics of BPS solutions.  Moreover, this also ensures that the solution to the bottom-up STU model is also a solution to the $\alpha'$-corrected heterotic theory.

At the two-derivative level, we can truncate to the pure supergravity theory by setting $A^1=A^2=A^3=A$ and $X^1=X^2=X^3=1$.  This truncation remains consistent at the four-derivative level provided we take identical couplings, $\lambda_1=\lambda_2=\lambda_3=\lambda/3$.  In this case, the four-derivative Lagrangian, (\ref{eq:4lagpart}), takes the form
\begin{align}
    e^{-1}\mathcal L&=R-\fft34F_{\mu\nu}^2+\fft14\epsilon^{\mu\nu\rho\sigma\lambda}F_{\mu\nu}F_{\rho\sigma}A_\lambda\nn\\
    &\quad+\lambda\left[R_{\mu\nu\rho\sigma}^2-\fft32R_{\mu\nu\rho\sigma}F^{\mu\nu}F^{\rho\sigma}+\fft54(F_{\mu\nu}^2)^2-\fft{39}8F^4+\fft12\epsilon^{\mu\nu\rho\sigma\lambda}R_{\mu\nu\alpha\beta}R_{\rho\sigma}{}^{\alpha\beta}A_\lambda\right],
\end{align}
and the corresponding four-derivative corrected BMPV black hole solution is
\begin{align}
    \dd s^2&=-e^{2f}(\dd t+\omega)^2+e^{2g}\dd r^2+e^{2h}\fft{r^2}4\left((\sigma_L^1)^2+(\sigma_L^2)^2+(\sigma_L^3)^2\right),\nn\\
    A&= \dfrac{1}{H}\Biggl[\left(1 + 2\fft\lambda{H}\left((\partial \log H)^2-\fft{14a^2}{3r^8H^3} \right)\right) (\dd t+\omega)+22\fft{\lambda}{H}\left(\fft{\partial\log H}r\right)\omega\Biggr],
\label{eq:bmpvm1}
\end{align}
with
\begin{align}
    e^{2f}&=\fft1{H^2}\qty(1-2\fft{\lambda}{H}\left(\fft{5(\partial\log H)^2}3+\fft{24a^2}{r^8H^3}\right)),\nn\\
    e^{2g}&=H\left(1-16\fft{\lambda}{H}\left(\fft{(\partial \log H)^2}3-\fft{a^2}{r^8H^3}\right)\right),\nn\\
    e^{2h}&=H\left(1+4\fft{\lambda}{H}\left(\fft{2(\partial\log H)^2}3+\fft{4a^2}{r^8H^3}\right)\right),\nn\\
    \omega&=\fft{a}{2r^2}\left(1-24\fft{\lambda}{H}\fft{\partial\log H}{r}\right)\sigma_L^3.
\end{align}

We can also perform a field redefinition
\begin{align}
    g_{\mu\nu}\to g_{\mu\nu}-\lambda\left(10F_{\mu\alpha}F_\nu{}^\alpha-\fft{11}6g_{\mu\nu}F^2\right),
\label{eq:fredef}
\end{align}
so that the four-derivative couplings take on a more streamlined form
\begin{align}
    e^{-1}\mathcal L&=R-\fft34F_{\mu\nu}^2+\fft14\epsilon^{\mu\nu\rho\sigma\lambda}F_{\mu\nu}F_{\rho\sigma}A_\lambda\nn\\
    &\quad+\lambda\left[\chi_{\mathrm{GB}}-\fft32C_{\mu\nu\rho\sigma}F^{\mu\nu}F^{\rho\sigma}+\fft98F^4+\fft12\epsilon^{\mu\nu\rho\sigma\lambda}R_{\mu\nu\alpha\beta}R_{\rho\sigma}{}^{\alpha\beta}A_\lambda\right].
\end{align}
In this frame, the black hole takes the form
\begin{align}
    \dd s^2&=-e^{2f}(\dd t+\omega)^2+e^{2g}\dd r^2+e^{2h}\fft{r^2}4\left((\sigma_L^1)^2+(\sigma_L^2)^2+(\sigma_L^3)^2\right),\nn\\
    A&= \dfrac{1}{H}\Biggl[\left(1 + 2\fft\lambda{H}\left((\partial \log H)^2-\fft{14a^2}{3r^8H^3} \right)\right) (\dd t+\omega)+2\fft{\lambda}{H}\left(\fft{\partial\log H}r\right)\omega\Biggr],
\end{align}
with
\begin{align}
    e^{2f}&=\fft1{H^2}\qty(1+\fft{\lambda}{H}\left(3(\partial\log H)^2-\fft{56a^2}{3r^8H^3}\right)),\nn\\
    e^{2g}&=H\left(1+\fft{\lambda}{H}\left((\partial \log H)^2+\fft{16a^2}{3r^8H^3}\right)\right),\nn\\
    e^{2h}&=H\left(1+\fft{\lambda}{H}\left(-(\partial\log H)^2+\fft{16a^2}{3r^8H^3}\right)\right),\nn\\
    \omega&=\fft{a}{2r^2}\left(1-4\fft{\lambda}{H}\fft{\partial\log H}{r}\right)\sigma_L^3.
\label{eq:bmpvm2}
\end{align}
Note that the graviphoton is not shifted by the field redefinition; the reason for the difference in the factor multiplying the last terms in the graviphoton expressions in (\ref{eq:bmpvm1}) and (\ref{eq:bmpvm2}) is that $\omega$ is shifted by (\ref{eq:fredef}).

\section{Higher-derivative corrections to the entropy}

Given the explicit BMPV solution, we now turn to the higher-derivative corrected entropy, which can be obtained using the Wald formula \cite{Wald:1993nt}
\begin{align}
    S = -2\pi \int_{\mathcal{H}} \varepsilon_{\mu\nu} \varepsilon_{\rho\sigma} \dfrac{\partial \mathcal{L}}{\partial R_{\mu\nu\rho\sigma}}\, \dd\Omega_3, \label{eq: wald entropy}
\end{align}
where the integral is evaluated over a cross section of the horizon $\mathcal{H}$ and $\epsilon_{\mu\nu}$ is the unit binormal to this slice. For two unit vectors $u,v$ normal to $\mathcal{H}$, the binormal is defined as,
\begin{align}
    \varepsilon = u \wedge v.
\end{align}
We will consider a constant time slice of the horizon. The unit normals to this slice can be chosen to be,
\begin{align}
    u = \dfrac{1}{\sqrt{1-w^2}} \left(e^f(\dd t + \omega) - \dfrac{w\, r}{2}e^h \sigma_L^3\right), \qquad v = e^g \,\dd r,
\end{align}
where we have defined,
\begin{align}
    w = \frac{a}{\sqrt{q_1q_2q_3}}.
\end{align}
The four-derivative part of the entropy will have two contributions, one arising from the corrections to the near-horizon geometry, which correct the horizon area, and the other from the four-derivative action. 

The horizon of the black hole is located at $r = 0$, which describes an $S^3$ of radius $R_H$. As noted in \cite{Cai:2025yyv} for the static case, the four-derivative corrections to the metric remain finite for the BMPV black hole as $r \to 0$. As a consequence of this, the near-horizon geometry remains $S^3$ fibered over $\mathrm{AdS}_2$. However, the radius $R_H$ of this $S^3$ receives corrections given by,
\begin{align}
    R_H = \sqrt{q_1q_2q_3 (1 - w^2)}\left[1 + \sum_I \dfrac{4\lambda_I}{q_I} \left(\dfrac{2w^4-7w^2 + 4 }{1-w^2}\right) \right].\label{eq: wald horizon part}
\end{align}
The second contribution to the entropy comes from the variation of the action with respect to the Riemann tensor. From the form of the four-derivative action, \eqref{eq:4lagpart}, we obtain
\begin{align}
    \dfrac{\partial \mathcal{L}}{\partial R_{\mu\nu\rho\sigma}} &= 2 \qty(g^{\mu\rho} g^{\nu\sigma} - g^{\mu\sigma} g^{\nu\rho}) \nn\\
    &\quad+ 8 \left(\lambda_IX^I R^{\mu\nu\rho\sigma} + \dfrac{1}{2} D_{IJ} F^{I \mu\nu} F^{J\rho\sigma} + \dfrac{1}{2} \lambda_I \epsilon^{\mu\nu\alpha\beta\gamma} R_{\alpha\beta}^{\quad \rho\sigma} A^I_\gamma\right).
\end{align}
For the BMPV black hole in \eqref{eq: BMPV alpha'}, one can compute this to be
\begin{equation}
    \varepsilon_{\mu\nu} \varepsilon_{\rho\sigma} \dfrac{\partial \mathcal{L}}{\partial R_{\mu\nu\rho\sigma}} = -8 + \sum_I\dfrac{32 \lambda_I}{q_I} (1 - 2w^2).\label{eq: wald 4der part}
\end{equation}
Combining \eqref{eq: wald horizon part} and \eqref{eq: wald 4der part} in \eqref{eq: wald entropy} and restoring the $G_N$ factors, we obtain the entropy of the BMPV black hole to be
\begin{equation}
    S =\dfrac{\pi^2}{2G_N} \sqrt{q_1q_2q_3 (1 - w^2)} \left(1+\sum_I\dfrac{12 \lambda_I}{q_I} \left(\dfrac{1 - 4w^2/3}{1-w^2}\right)\right).\label{eq:entropy}
\end{equation}
This agrees exactly with the results of \cite{Ruiperez:2026rni, Cassani:2024tvk} and \cite{Alexandrov:2026rra}.  Note that, since we only work to first order in the $\lambda_I$ couplings, the entropy can be rewritten as
\begin{equation}
    S=\dfrac{\pi^2}{2G_N} \sqrt{(q_1+24\lambda_1g(w))(q_2+24\lambda_2g(w))(q_3+24\lambda_3g(w))(1-w^2)},
\end{equation}
where
\begin{equation}
    g(w)=\fft{1-4w^2/3}{1-w^2}.
\end{equation}
When $w=0$, this reduces to the familiar expression for the higher-derivative corrected static STU black hole entropy \cite{Castro:2007hc,Castro:2008ys,DominisPrester:2008ynb,Faedo:2019xii,Elgood:2020xwu,Cano:2021nzo,Cai:2025yyv,Cai:2026qrr}
\begin{equation}
    S=\dfrac{\pi^2}{2G_N} \sqrt{(q_1+24\lambda_1)(q_2+24\lambda_2)(q_3+24\lambda_3)}.
\end{equation}
%

\section{Conclusions}

In this paper, we have used AI to construct the four-derivative corrected solution for the BMPV black hole directly in the five-dimensional STU model with the standard Weyl four-derivative superinvariant \cite{Hanaki:2006pj,Cassani:2024tvk}.  As shown in \cite{Cai:2026qrr}, this model differs from that obtained by dimensional reduction of the $\alpha'$-corrected heterotic theory.  However, the difference is due to vector multiplet invariants which do not modify the BMPV black hole solution.  Thus, the solution that we obtained in five dimensions should also be a solution to the heterotic theory, provided we set $\lambda_1=\lambda_2=0$ and $\lambda_3=\alpha'/8$.  In particular, the solution, (\ref{eq: BMPV alpha'}), should be equivalent to that found in~\cite{Ruiperez:2026rni}. A comparison would require reducing the ten-dimensional solution on $T^5$ and then performing appropriate field redefinitions~\cite{Jayaprakash:2024xlr,Cai:2025yyv}. As we have directly constructed and verified the solution in five dimensions, we have not checked that the two solutions are equivalent after field redefinitions, although consistency dictates that they should be.

While the higher-derivative corrected entropy can be obtained without explicit knowledge of the four-derivative corrected solution, we can nevertheless gain confidence in the result by performing a direct computation of the Wald entropy from the corrected solution.  The five-dimensional Wald entropy computation parallels that of the ten-dimensional heterotic case \cite{Ruiperez:2026rni}, and yields the same result. Nevertheless, as discussed in \cite{Ruiperez:2026rni,Alexandrov:2026rra}, this entropy differs from that of some previous calculations, including those in~\cite{Guica:2005ig,Castro:2007ci,Castro:2008ys,deWit:2009de,Gupta:2021roy}.  Given that the entropy obtained here and in Ref.~\cite{Ruiperez:2026rni} was computed directly from the full $\alpha'$-corrected BMPV solution, we have reason to believe that this result is valid.  However, some subtleties may arise in obtaining the Wald entropy in the presence of the Chern-Simons term $A\wedge\Tr R\wedge R$ \cite{Tachikawa:2006sz}. The different results obtained in earlier literature also stem from the fact that the conserved charges are extracted from the near-horizon geometry rather than from the full asymptotically flat solution. Hence, further investigation is needed to fully resolve the nature of these discrepancies.

It is interesting to note that the four-derivative corrections to the entropy~\eqref{eq:entropy} become negative for $3/4<w^2<1$. This is in contradiction to the expectation from the weak gravity conjecture~\cite{Cheung:2018cwt}; however, there are several other non-supersymmetric counterexamples that have been found previously~\cite{Ma:2022gtm,Wu:2024iiz}.

The GPT 5.6 calculation and verification material can be obtained from GitHub at the link \href{https://github.com/wuhucyi/bmpv-higher-derivative-verification/tree/main}{https://github.com/wuhucyi/bmpv-higher-derivative-verification/tree/main}.  This material is organized for both human consumption and AI agents following the Agent-Native Research Artifact protocol \cite{liu2026humanwrittenpaperagentnativeresearch}.

Finally, we believe the problem of finding black hole solutions is ideally suited for AI, as it is a well-posed problem, we expect a well-defined solution, and we can check the candidate solution directly. While solutions with rotation are generally quite difficult to solve exactly by direct computation, as a BPS solution in ungauged supergravity, the two-derivative BMPV black hole provides a relatively simple starting point for constructing the higher-derivative corrections.  It would be exciting to see how far AI can go in working out more challenging classes of solutions.

\section*{Acknowledgments}

This material is based upon work supported by the U.S.~Department of Energy, Office of Science, Office of High Energy Physics, under Award Number DE-SC0026542. The work of YP and RJS is supported by the National Natural Science Foundation of China (NSFC)
under Grants No.~12575076 and No.~12247103.

\noindent\textit{Disclaimer.} This report was prepared as an account of work sponsored by an agency of the United States Government.  Neither the United States Government nor any agency thereof, nor any of their employees, makes any warranty, express or implied, or assumes any legal liability or responsibility for the accuracy, completeness, or usefulness of any information, apparatus, product, or process disclosed, or represents that its use would not infringe privately owned rights.  Reference herein to any specific commercial product, process, or service by trade name, trademark, manufacturer, or otherwise does not necessarily constitute or imply its endorsement, recommendation, or favoring by the United States Government or any agency thereof.  The views and opinions of authors expressed herein do not necessarily state or reflect those of the United States Government or any agency thereof.

\bibliographystyle{JHEP}
\bibliography{biblio.bib}

\providecommand{\href}[2]{#2}\begingroup\raggedright\begin{thebibliography}{10}

\bibitem{Strominger:1996sh}
A.~Strominger and C.~Vafa, \emph{{Microscopic origin of the Bekenstein-Hawking entropy}}, \href{https://doi.org/10.1016/0370-2693(96)00345-0}{\emph{Phys. Lett. B} {\bfseries 379} (1996) 99} [\href{https://arxiv.org/abs/hep-th/9601029}{{\ttfamily hep-th/9601029}}].

\bibitem{Breckenridge:1996is}
J.C.~Breckenridge, R.C.~Myers, A.W.~Peet and C.~Vafa, \emph{{D-branes and spinning black holes}}, \href{https://doi.org/10.1016/S0370-2693(96)01460-8}{\emph{Phys. Lett. B} {\bfseries 391} (1997) 93} [\href{https://arxiv.org/abs/hep-th/9602065}{{\ttfamily hep-th/9602065}}].

\bibitem{Maldacena:1997de}
J.M.~Maldacena, A.~Strominger and E.~Witten, \emph{{Black hole entropy in M theory}}, \href{https://doi.org/10.1088/1126-6708/1997/12/002}{\emph{JHEP} {\bfseries 12} (1997) 002} [\href{https://arxiv.org/abs/hep-th/9711053}{{\ttfamily hep-th/9711053}}].

\bibitem{Bergshoeff:1988nn}
E.~Bergshoeff and M.~de~Roo, \emph{{Supersymmetric Chern-simons Terms in Ten-dimensions}}, \href{https://doi.org/10.1016/0370-2693(89)91420-2}{\emph{Phys. Lett. B} {\bfseries 218} (1989) 210}.

\bibitem{Bergshoeff:1989de}
E.A.~Bergshoeff and M.~de~Roo, \emph{{The Quartic Effective Action of the Heterotic String and Supersymmetry}}, \href{https://doi.org/10.1016/0550-3213(89)90336-2}{\emph{Nucl. Phys. B} {\bfseries 328} (1989) 439}.

\bibitem{Reall:2019sah}
H.S.~Reall and J.E.~Santos, \emph{{Higher derivative corrections to Kerr black hole thermodynamics}}, \href{https://doi.org/10.1007/JHEP04(2019)021}{\emph{JHEP} {\bfseries 04} (2019) 021} [\href{https://arxiv.org/abs/1901.11535}{{\ttfamily 1901.11535}}].

\bibitem{Ruiperez:2026rni}
A.~Ruip{\'e}rez, \emph{{BMPV black hole at first order in $\alpha'$}},  \href{https://arxiv.org/abs/2606.31987}{{\ttfamily 2606.31987}}.

\bibitem{Jayaprakash:2024xlr}
S.~Jayaprakash and J.T.~Liu, \emph{{Higher derivative heterotic supergravity on a torus and supersymmetry}}, \href{https://doi.org/10.1007/JHEP12(2024)076}{\emph{JHEP} {\bfseries 12} (2024) 076} [\href{https://arxiv.org/abs/2406.14600}{{\ttfamily 2406.14600}}].

\bibitem{Cai:2025yyv}
Y.~Cai, S.~Jayaprakash, J.T.~Liu and R.J.~Saskowski, \emph{{Multicenter higher-derivative BPS black holes}}, \href{https://doi.org/10.1007/JHEP10(2025)014}{\emph{JHEP} {\bfseries 10} (2025) 014} [\href{https://arxiv.org/abs/2502.05065}{{\ttfamily 2502.05065}}].

\bibitem{Cai:2026qrr}
Y.~Cai, S.~Jayaprakash, J.T.~Liu, Y.~Pang and R.J.~Saskowski, \emph{{New F$^{4}$ invariants in five-dimensional supergravity}}, \href{https://doi.org/10.1007/JHEP07(2026)065}{\emph{JHEP} {\bfseries 07} (2026) 065} [\href{https://arxiv.org/abs/2603.18180}{{\ttfamily 2603.18180}}].

\bibitem{Hanaki:2006pj}
K.~Hanaki, K.~Ohashi and Y.~Tachikawa, \emph{{Supersymmetric Completion of an R**2 term in Five-dimensional Supergravity}}, \href{https://doi.org/10.1143/PTP.117.533}{\emph{Prog. Theor. Phys.} {\bfseries 117} (2007) 533} [\href{https://arxiv.org/abs/hep-th/0611329}{{\ttfamily hep-th/0611329}}].

\bibitem{Chang:2022urm}
H.-Y.~Chang, E.~Sezgin and Y.~Tanii, \emph{{Dualization of higher derivative heterotic supergravities in 6D and 10D}}, \href{https://doi.org/10.1007/JHEP10(2022)062}{\emph{JHEP} {\bfseries 10} (2022) 062} [\href{https://arxiv.org/abs/2209.03981}{{\ttfamily 2209.03981}}].

\bibitem{Bergshoeff:1986vy}
E.~Bergshoeff, A.~Salam and E.~Sezgin, \emph{{A Supersymmetric R**2 Action in Six-dimensions and Torsion}}, \href{https://doi.org/10.1016/0370-2693(86)91233-5}{\emph{Phys. Lett. B} {\bfseries 173} (1986) 73}.

\bibitem{Novak:2017wqc}
J.~Novak, M.~Ozkan, Y.~Pang and G.~Tartaglino-Mazzucchelli, \emph{{Gauss-Bonnet supergravity in six dimensions}}, \href{https://doi.org/10.1103/PhysRevLett.119.111602}{\emph{Phys. Rev. Lett.} {\bfseries 119} (2017) 111602} [\href{https://arxiv.org/abs/1706.09330}{{\ttfamily 1706.09330}}].

\bibitem{Cassani:2024tvk}
D.~Cassani, A.~Ruip{\'e}rez and E.~Turetta, \emph{{Higher-derivative corrections to flavoured BPS black hole thermodynamics and holography}}, \href{https://doi.org/10.1007/JHEP05(2024)276}{\emph{JHEP} {\bfseries 05} (2024) 276} [\href{https://arxiv.org/abs/2403.02410}{{\ttfamily 2403.02410}}].

\bibitem{Wald:1993nt}
R.M.~Wald, \emph{{Black hole entropy is the Noether charge}}, \href{https://doi.org/10.1103/PhysRevD.48.R3427}{\emph{Phys. Rev. D} {\bfseries 48} (1993) R3427} [\href{https://arxiv.org/abs/gr-qc/9307038}{{\ttfamily gr-qc/9307038}}].

\bibitem{Alexandrov:2026rra}
S.~Alexandrov, A.~Klemm and B.~Pioline, \emph{{Large Order Enumerative Geometry, Black Holes and Black Rings}},  \href{https://arxiv.org/abs/2605.19552}{{\ttfamily 2605.19552}}.

\bibitem{Castro:2007hc}
A.~Castro, J.L.~Davis, P.~Kraus and F.~Larsen, \emph{{5D Black Holes and Strings with Higher Derivatives}}, \href{https://doi.org/10.1088/1126-6708/2007/06/007}{\emph{JHEP} {\bfseries 06} (2007) 007} [\href{https://arxiv.org/abs/hep-th/0703087}{{\ttfamily hep-th/0703087}}].

\bibitem{Castro:2008ys}
A.~Castro and S.~Murthy, \emph{{Corrections to the statistical entropy of five dimensional black holes}}, \href{https://doi.org/10.1088/1126-6708/2009/06/024}{\emph{JHEP} {\bfseries 06} (2009) 024} [\href{https://arxiv.org/abs/0807.0237}{{\ttfamily 0807.0237}}].

\bibitem{DominisPrester:2008ynb}
P.~Dominis~Prester and T.~Terzic, \emph{{$\alpha'$-exact entropies for BPS and non-BPS extremal dyonic black holes in heterotic string theory from ten-dimensional supersymmetry}}, \href{https://doi.org/10.1088/1126-6708/2008/12/088}{\emph{JHEP} {\bfseries 12} (2008) 088} [\href{https://arxiv.org/abs/0809.4954}{{\ttfamily 0809.4954}}].

\bibitem{Faedo:2019xii}
F.~Faedo and P.F.~Ramirez, \emph{{Exact charges from heterotic black holes}}, \href{https://doi.org/10.1007/JHEP10(2019)033}{\emph{JHEP} {\bfseries 10} (2019) 033} [\href{https://arxiv.org/abs/1906.12287}{{\ttfamily 1906.12287}}].

\bibitem{Elgood:2020xwu}
Z.~Elgood and T.~Ortin, \emph{{T duality and Wald entropy formula in the Heterotic Superstring effective action at first-order in {\ensuremath{\alpha}}'}}, \href{https://doi.org/10.1007/JHEP10(2020)097}{\emph{JHEP} {\bfseries 10} (2020) 097} [\href{https://arxiv.org/abs/2005.11272}{{\ttfamily 2005.11272}}].

\bibitem{Cano:2021nzo}
P.A.~Cano, T.~Ort{\'\i}n, A.~Ruip{\'e}rez and M.~Zatti, \emph{{Non-supersymmetric black holes with {\ensuremath{\alpha}}' corrections}}, \href{https://doi.org/10.1007/JHEP03(2022)103}{\emph{JHEP} {\bfseries 03} (2022) 103} [\href{https://arxiv.org/abs/2111.15579}{{\ttfamily 2111.15579}}].

\bibitem{Guica:2005ig}
M.~Guica, L.~Huang, W.~Li and A.~Strominger, \emph{{R**2 corrections for 5-D black holes and rings}}, \href{https://doi.org/10.1088/1126-6708/2006/10/036}{\emph{JHEP} {\bfseries 10} (2006) 036} [\href{https://arxiv.org/abs/hep-th/0505188}{{\ttfamily hep-th/0505188}}].

\bibitem{Castro:2007ci}
A.~Castro, J.L.~Davis, P.~Kraus and F.~Larsen, \emph{{Precision Entropy of Spinning Black Holes}}, \href{https://doi.org/10.1088/1126-6708/2007/09/003}{\emph{JHEP} {\bfseries 09} (2007) 003} [\href{https://arxiv.org/abs/0705.1847}{{\ttfamily 0705.1847}}].

\bibitem{deWit:2009de}
B.~de~Wit and S.~Katmadas, \emph{{Near-Horizon Analysis of D=5 BPS Black Holes and Rings}}, \href{https://doi.org/10.1007/JHEP02(2010)056}{\emph{JHEP} {\bfseries 02} (2010) 056} [\href{https://arxiv.org/abs/0910.4907}{{\ttfamily 0910.4907}}].

\bibitem{Gupta:2021roy}
R.K.~Gupta, S.~Murthy and M.~Sahni, \emph{{Quantum entropy of BMPV black holes and the topological M-theory conjecture}}, \href{https://doi.org/10.1007/JHEP06(2022)053}{\emph{JHEP} {\bfseries 06} (2022) 053} [\href{https://arxiv.org/abs/2104.02634}{{\ttfamily 2104.02634}}].

\bibitem{Tachikawa:2006sz}
Y.~Tachikawa, \emph{{Black hole entropy in the presence of Chern-Simons terms}}, \href{https://doi.org/10.1088/0264-9381/24/3/014}{\emph{Class. Quant. Grav.} {\bfseries 24} (2007) 737} [\href{https://arxiv.org/abs/hep-th/0611141}{{\ttfamily hep-th/0611141}}].

\bibitem{Cheung:2018cwt}
C.~Cheung, J.~Liu and G.N.~Remmen, \emph{{Proof of the Weak Gravity Conjecture from Black Hole Entropy}}, \href{https://doi.org/10.1007/JHEP10(2018)004}{\emph{JHEP} {\bfseries 10} (2018) 004} [\href{https://arxiv.org/abs/1801.08546}{{\ttfamily 1801.08546}}].

\bibitem{Ma:2022gtm}
L.~Ma, Y.~Pang and H.~L{\"u}, \emph{{Negative corrections to black hole entropy from string theory}}, \href{https://doi.org/10.1007/s11433-023-2257-6}{\emph{Sci. China Phys. Mech. Astron.} {\bfseries 66} (2023) 121011} [\href{https://arxiv.org/abs/2212.03262}{{\ttfamily 2212.03262}}].

\bibitem{Wu:2024iiz}
P.-Y.~Wu and H.~Lu, \emph{{Quadratic curvature correction and its breakdown to thermodynamics of rotating black holes}}, \href{https://doi.org/10.1103/PhysRevD.111.104026}{\emph{Phys. Rev. D} {\bfseries 111} (2025) 104026} [\href{https://arxiv.org/abs/2405.04576}{{\ttfamily 2405.04576}}].

\bibitem{liu2026humanwrittenpaperagentnativeresearch}
J.~Liu, J.~Pei, J.~Huang, C.~Si, A.~Qu, X.~Tang et~al., \emph{The last human-written paper: Agent-native research artifacts},  \href{https://arxiv.org/abs/2604.24658}{{\ttfamily 2604.24658}}.

\end{thebibliography}\endgroup

\end{document}